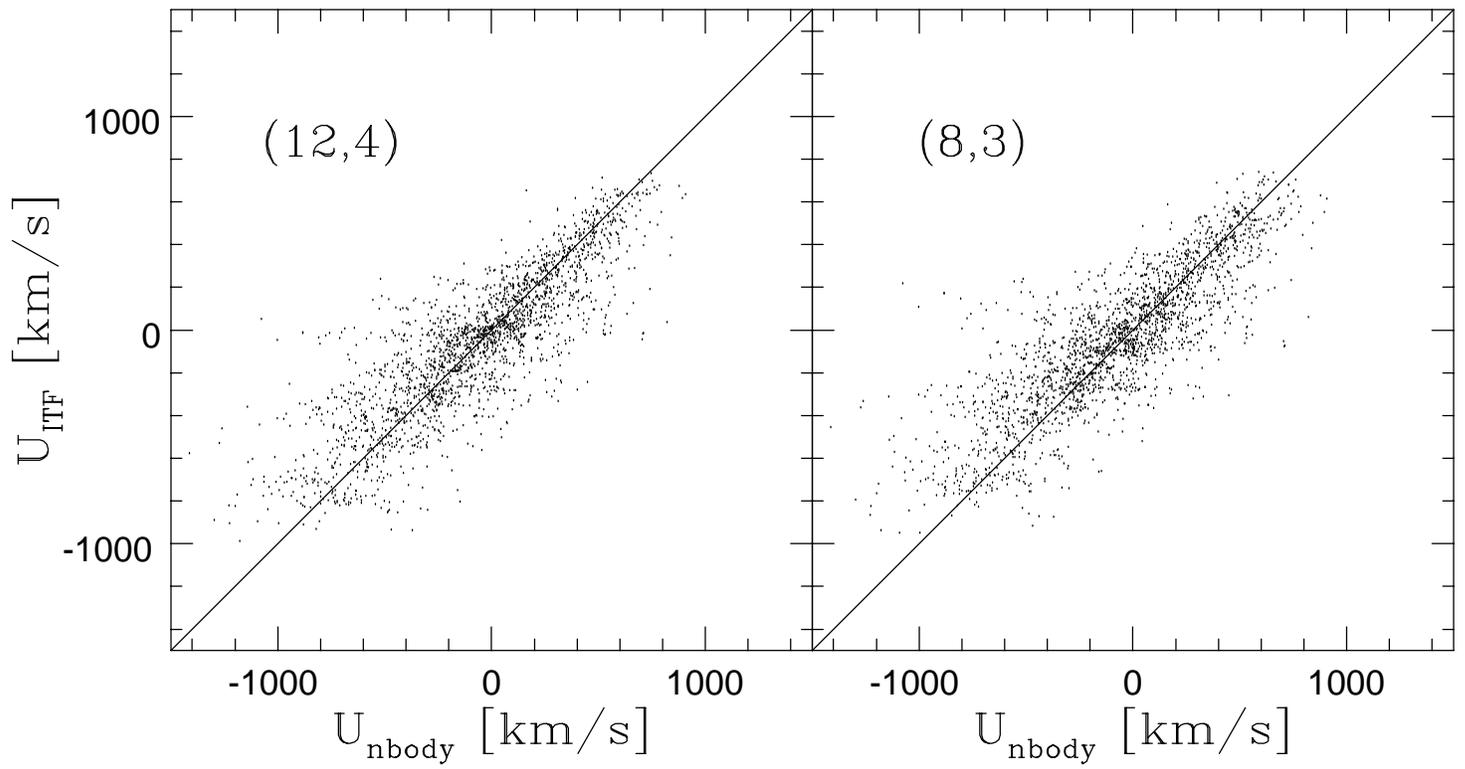

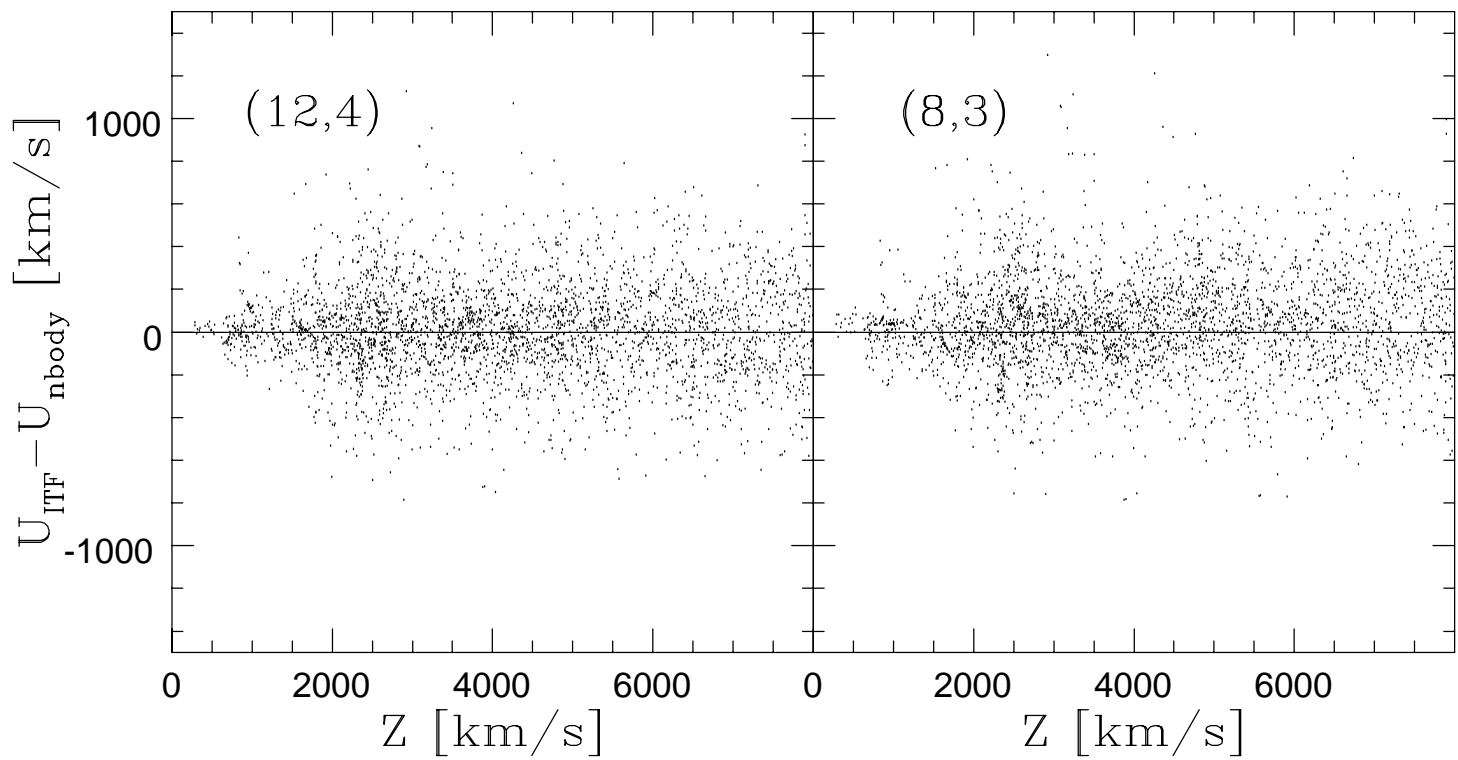

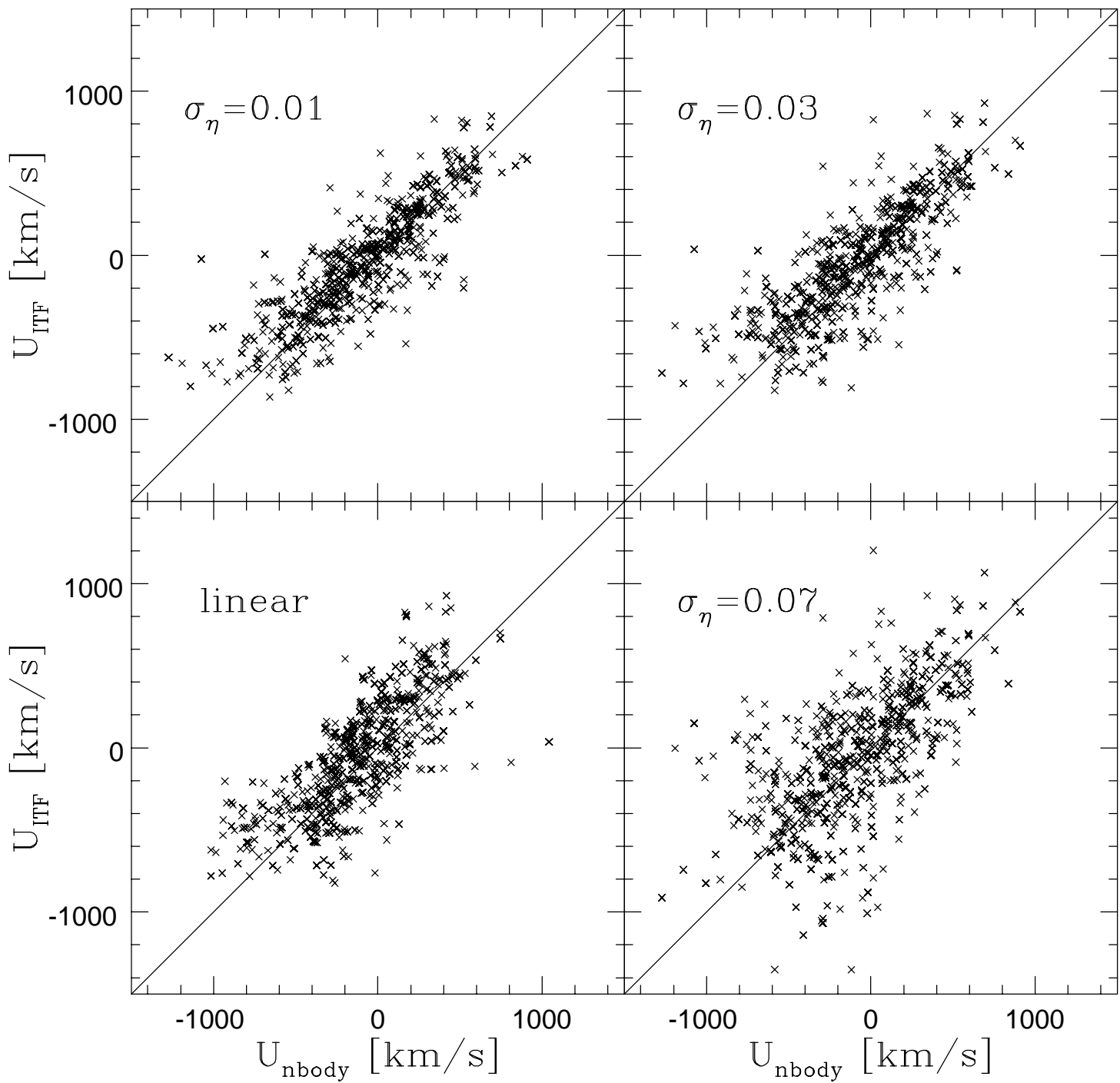

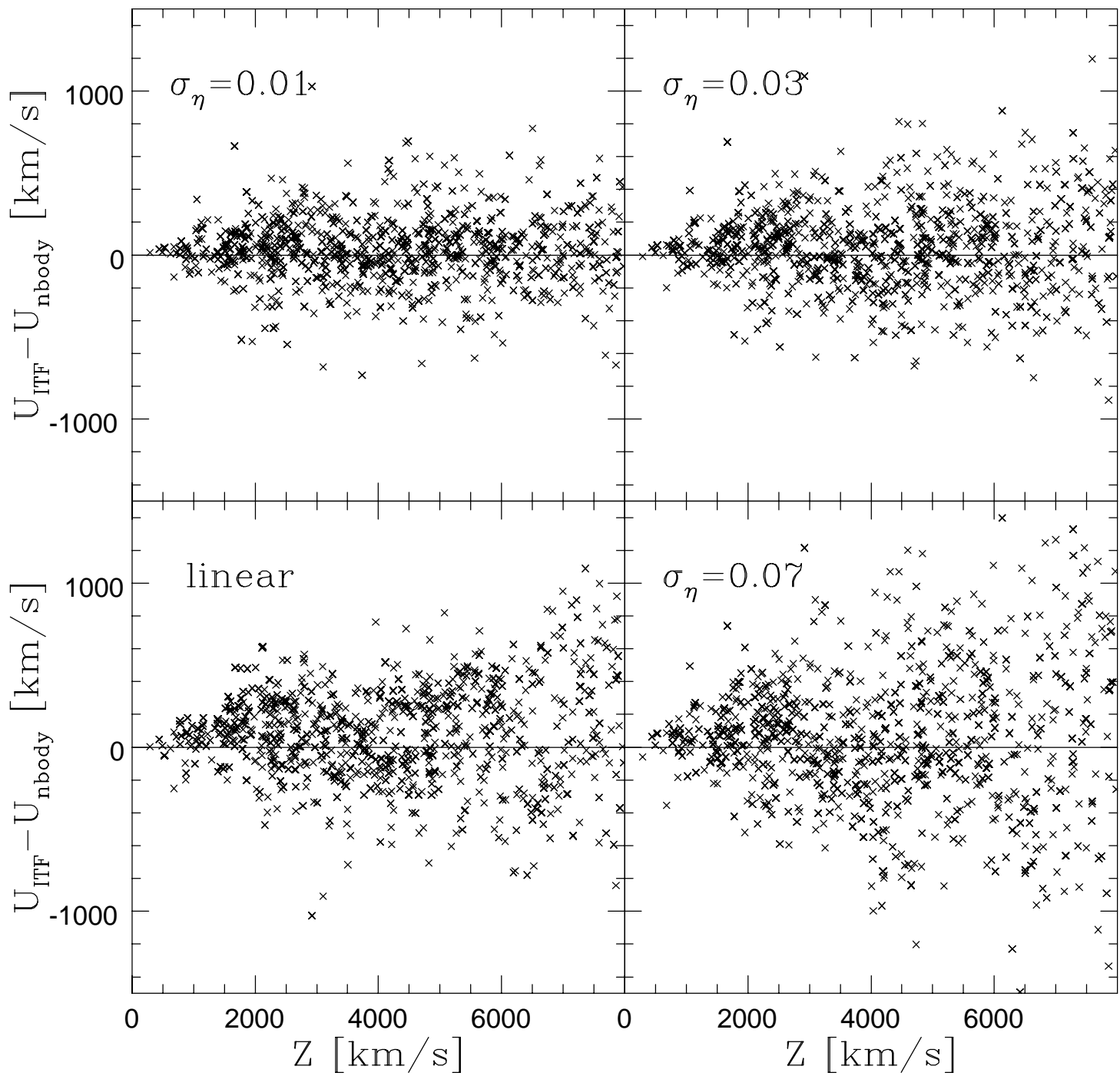

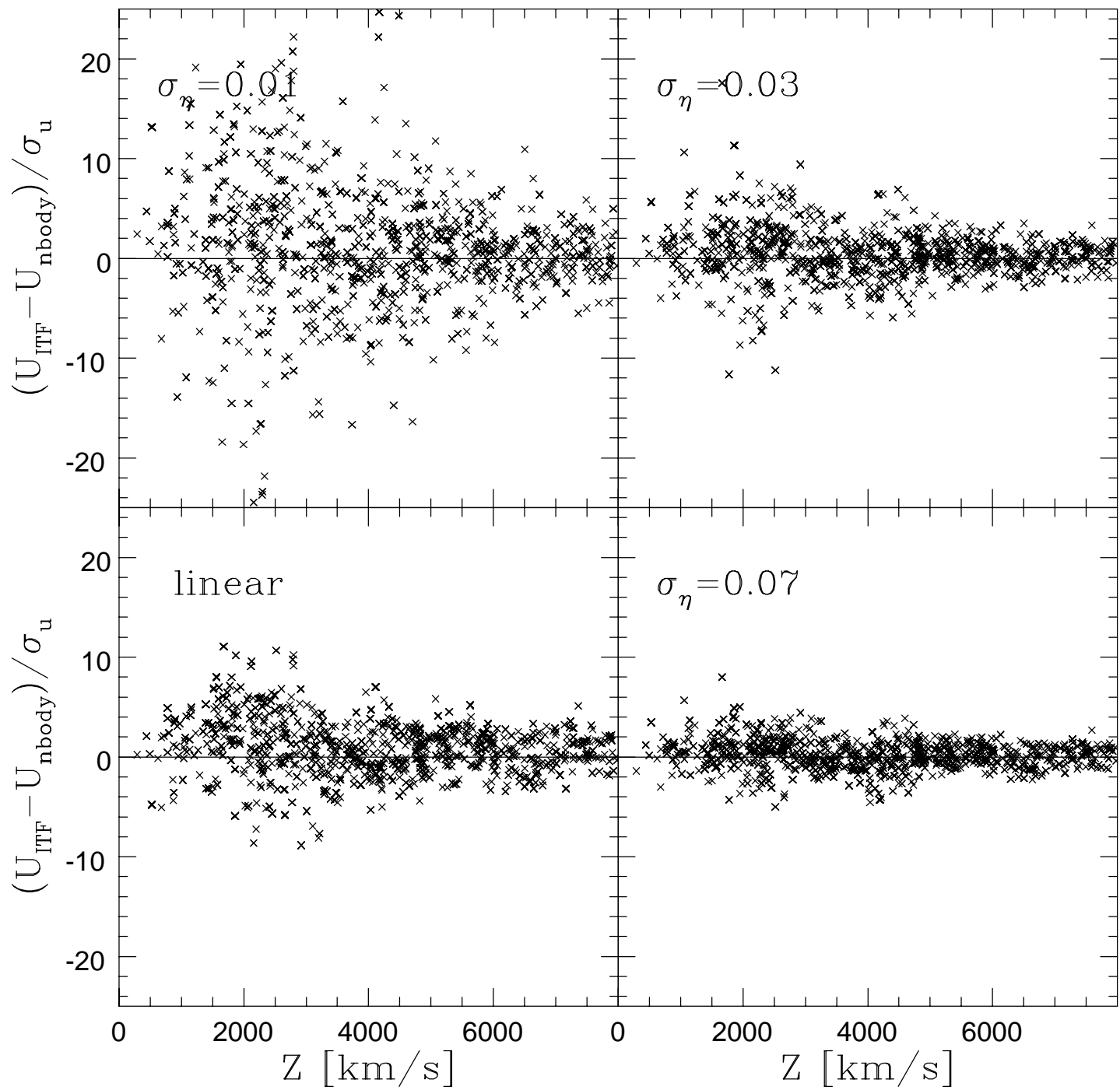

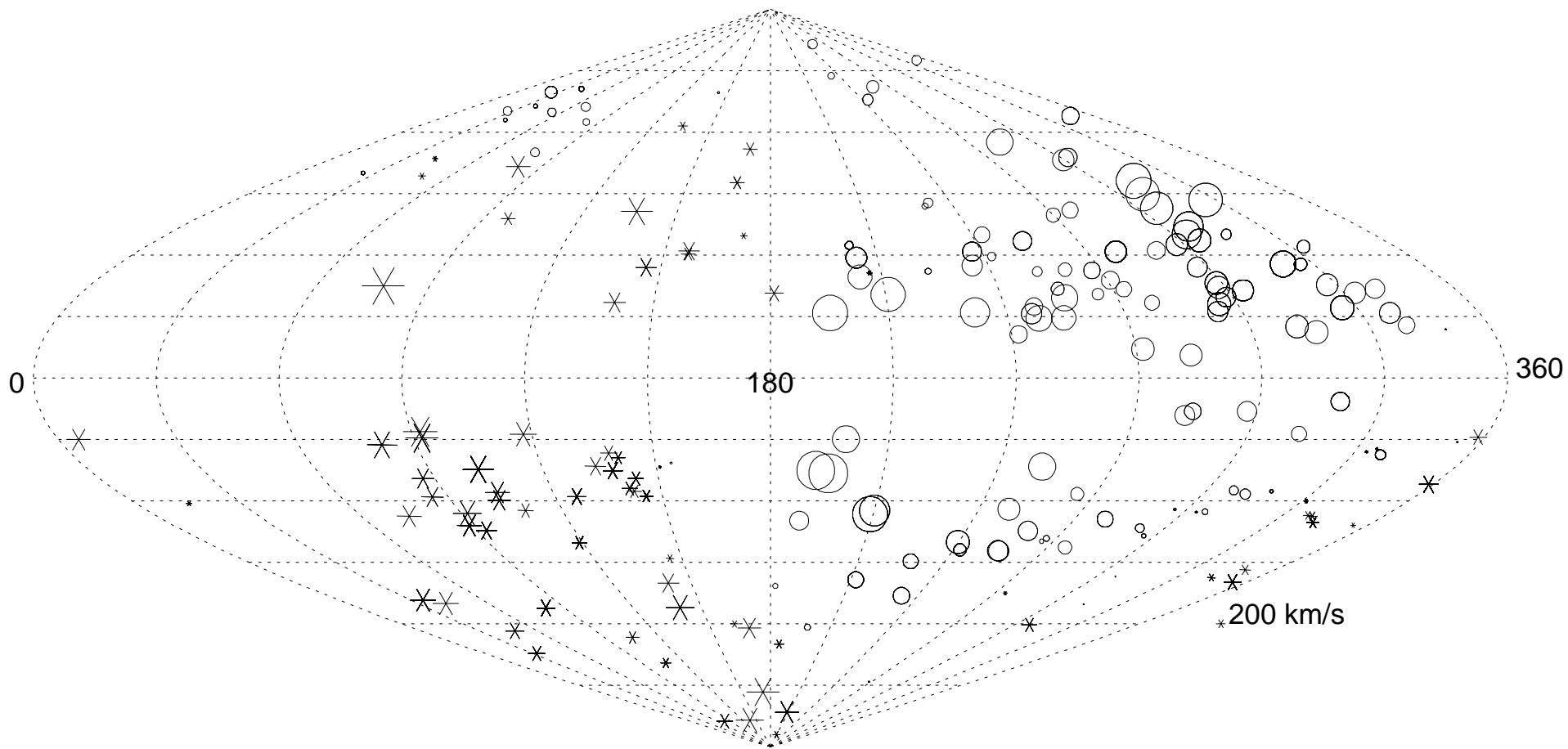

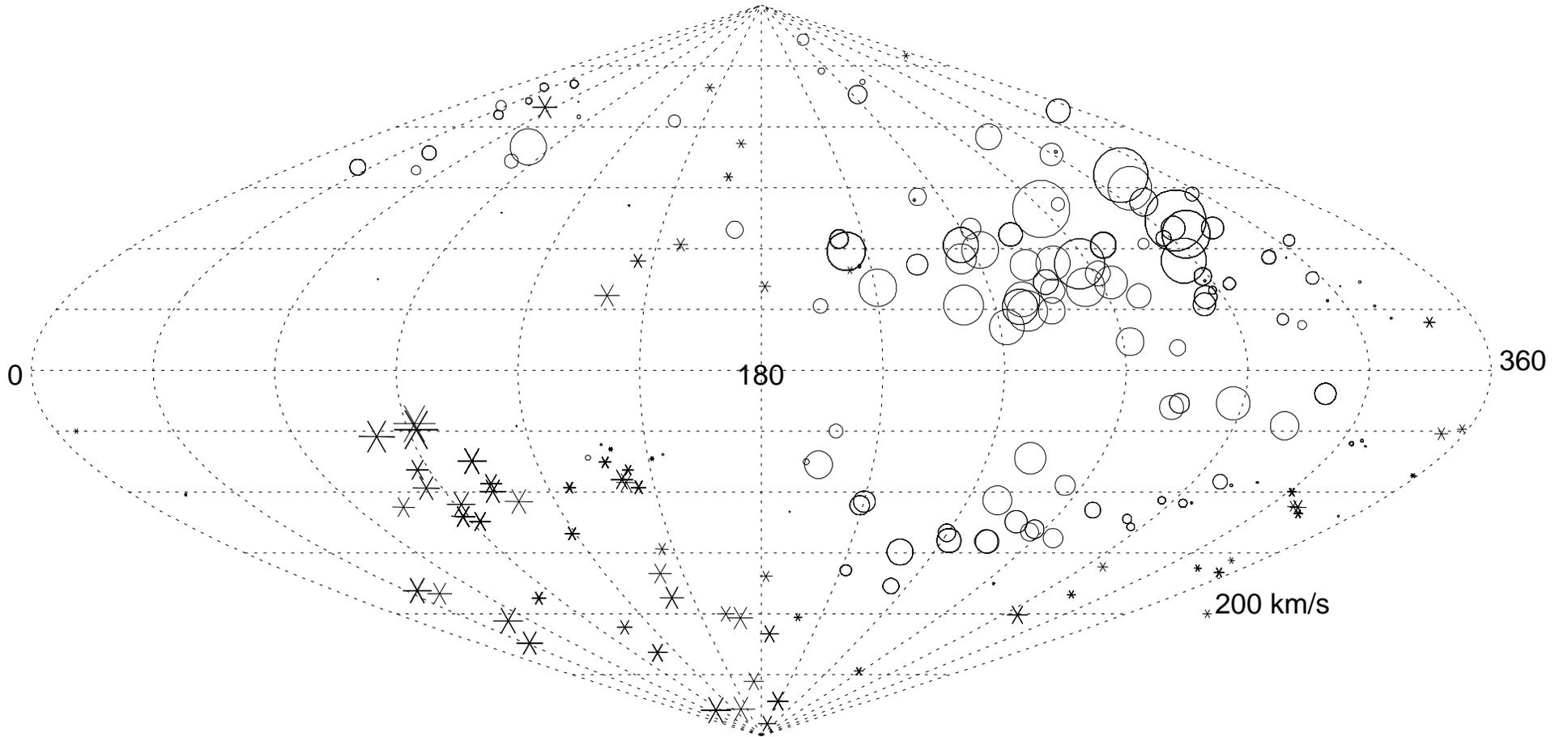

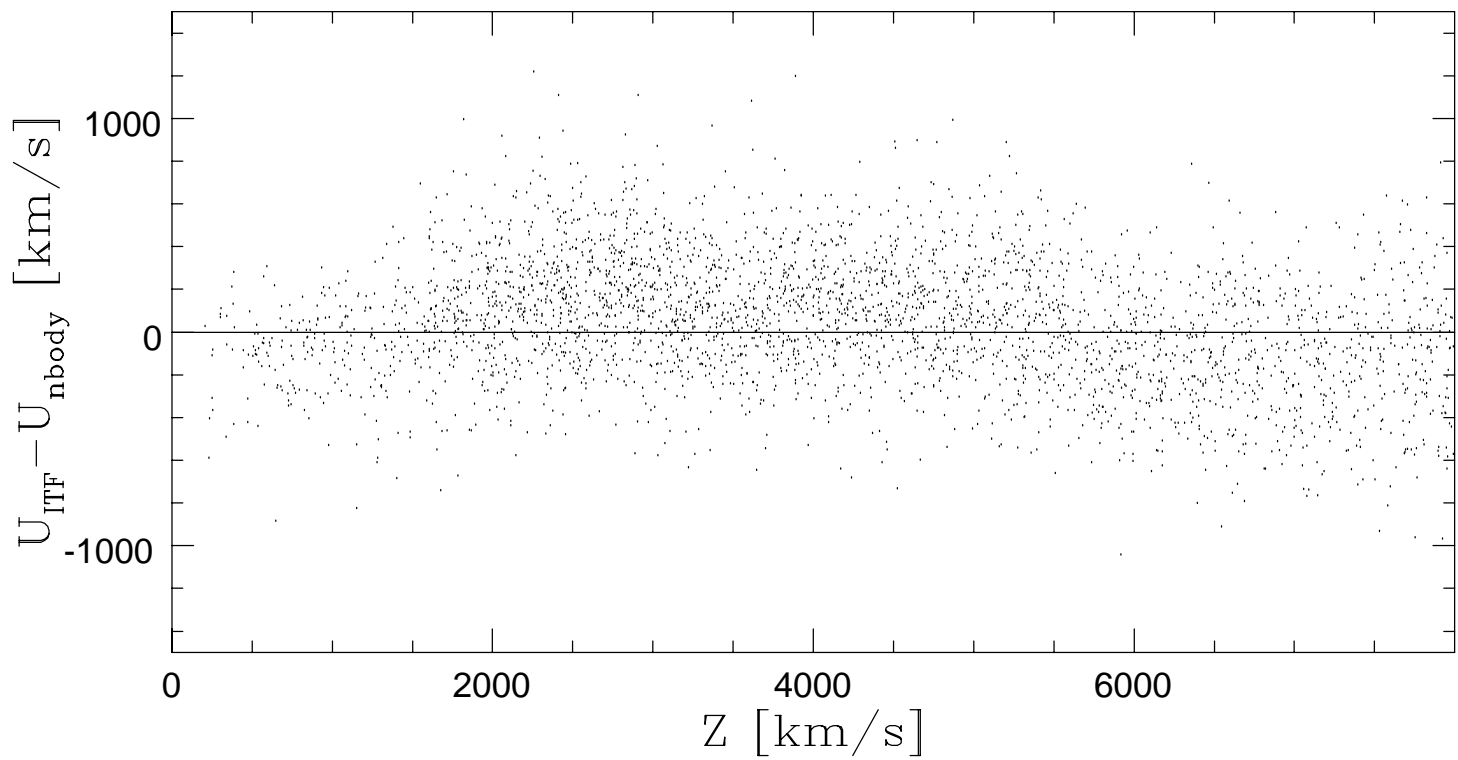

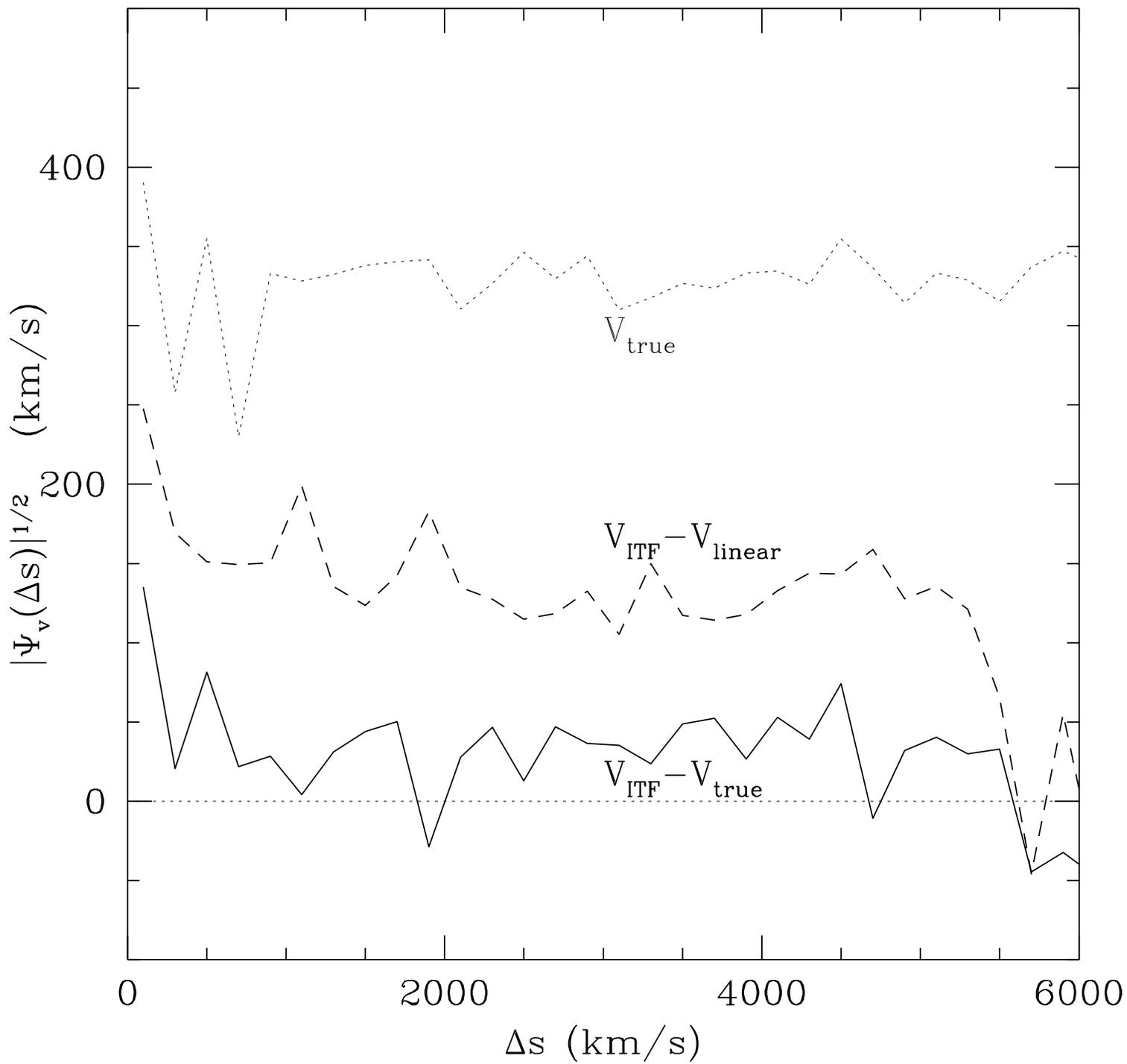

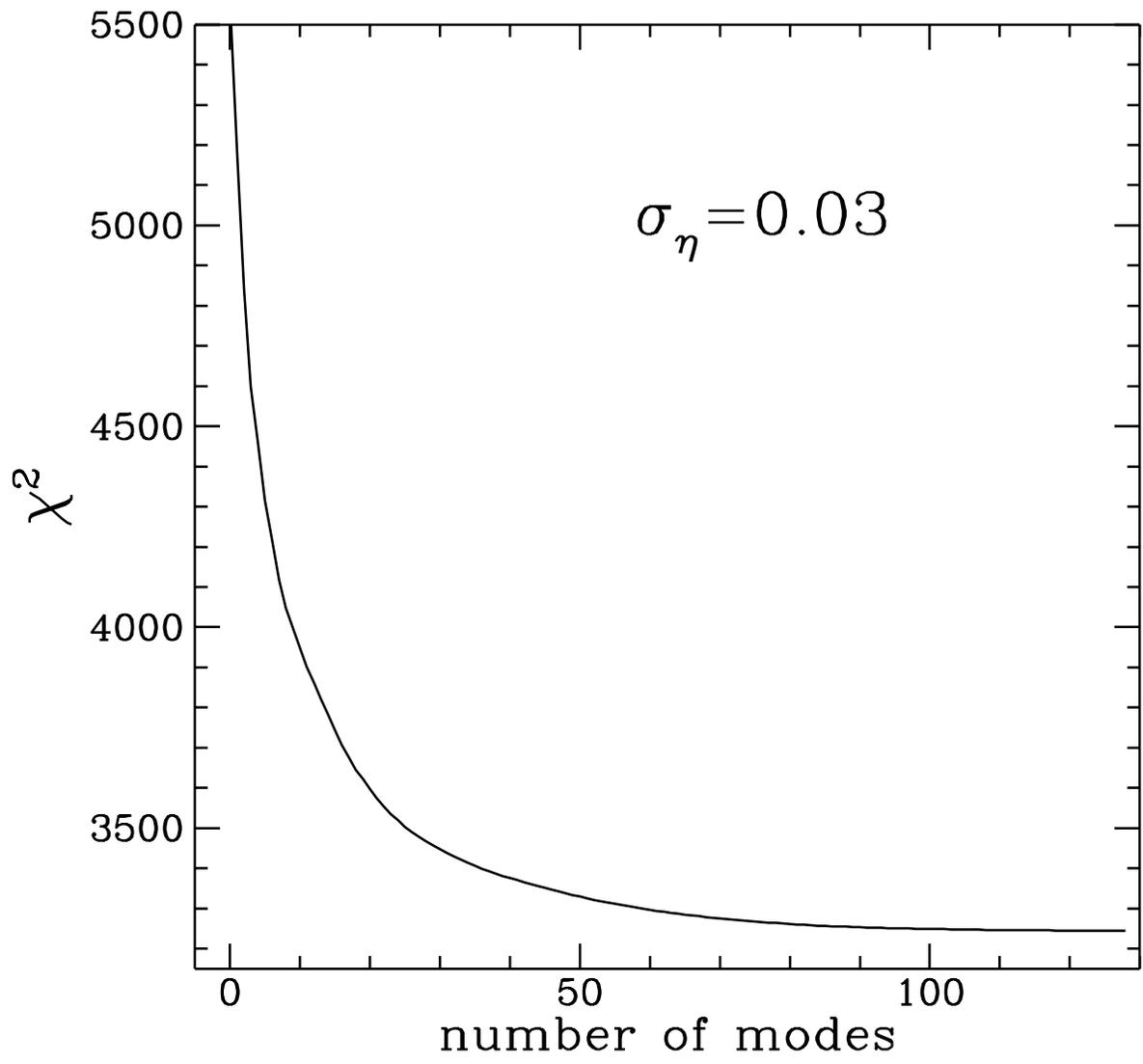

# ESTIMATION OF PECULIAR VELOCITY FROM THE INVERSE TULLY-FISHER RELATION

**Adi Nusser**
Institute of Astronomy, Madingley Rd, Cambridge, CB3 0HA, England
and
**Marc Davis**
Departments of Astronomy and Physics
University of California, Berkeley, CA. 94720, U.S.A

## ABSTRACT

We present a method for deriving a smoothed estimate of the peculiar velocity field of a set of galaxies with measured circular velocities $\eta \equiv \log\Delta v$ and apparent magnitudes $m$. The method is based on minimizing the scatter of a linear inverse Tully-Fisher relation $\eta = \eta(M)$ where the absolute magnitude of each galaxy is inferred from its redshift $z$, corrected by a peculiar velocity field, $M \propto m - 5\log(z - u)$. We describe the radial peculiar velocity field $u(\mathbf{z})$ in terms of a set of orthogonal functions which can be derived from any convenient basis set; as an example we take them to be linear combinations of low order spherical harmonic and spherical Bessel functions. The model parameters are then found by maximizing the likelihood function for measuring a set of observed $\eta$. The predicted peculiar velocities are free of Malmquist bias in the absence of multi-streaming, provided no selection criteria are imposed on the measurement of circular velocities. This procedure can be considered as a generalized smoothing algorithm of the peculiar velocity field, and is particularly useful for comparison to the smoothed gravity field derived from full-sky galaxy redshift catalogs such as the IRAS surveys. We demonstrate the technique using a catalog of "galaxies" derived from an N-body simulation. Increasing the resolution of the velocity smoothing beyond a certain level degrades the correlation of fitted velocities against the velocities calculated from linear theory methods, which have finite resolution, but the slope of the scatter diagram, and therefore the derived density parameter, remains fixed.

*Subject headings:* cosmology — dark matter — galaxies: clustering — galaxies: formation — gravity — large-scale structure of the universe

# 1. INTRODUCTION

The large scale motions of galaxies in the universe are an interesting probe of the nature of large scale structure in the Universe. Indeed, if galaxies move under the influence of the gravity of the underlying matter field, then their peculiar velocities, the deviations from smooth Hubble expansion, can be used to address fundamental questions such as the amplitude of the clustering in the dark matter distribution, the mean density of the Universe, and the statistical nature of the primordial fluctuations. A thorough review of this active field is given by Dekel (1994).

All of the methods of mapping deviations from Hubble flow depend on an observable parameter that is distance dependent, e.g. magnitude, and another which is distance independent, e.g. line width or rotational velocity. The remarkable Tully-Fisher (TF) correlation between rotational velocity and absolute magnitude for disklike galaxies is the most commonly used distance indicator, with a distance error per galaxy of order 16%. Alternative extragalactic distance estimators have been recently reviewed by Jacoby *et al.* (1992).

In a widely cited recent publication, Dekel *et al.* (1993) compared the density field of IRAS selected galaxies with the divergence of the measured peculiar velocity field derived from the POTENT algorithm (Bertschinger *et al.* 1990; Dekel, Bertschinger, and Faber 1990). They have shown that the flows are directed toward the peaks of the galaxy density, as expected in the linear theory of gravitational instability (Peebles 1980); quantitatively, they derive a high estimate of the density parameter, $\beta \equiv \Omega^{0.6}/b_I \approx 1.3 \pm 0.3$, where $b_I$ is the linear bias in the IRAS galaxy distribution relative to the mass distribution.

The Dekel *et al.* (1993) result depended on a forward fit (i.e. $M = M(\eta)$), of the Tully Fisher relation for spiral galaxies, and the equivalent $D_n - \sigma$ relationship for the elliptical galaxies, and is subject to homogeneous as well as inhomogeneous Malmquist bias (see e.g. Lynden-Bell *et al.* 1988, Gould 1993, Willick 1994). These effects must be carefully calibrated, and are quite sensitive to the estimated uncertainty in the TF scatter. The POTENT method requires extensive smoothing, as well as the gridding of the peculiar velocity data, which can introduce a serious "sampling gradient bias".

Such a complex analysis must be tested by a number of independent means, using the best available data. Willick *et al.* (1994b) are using the newly compiled Mark3 dataset (Willick *et al.* 1994a) of $\approx 2800$ spiral galaxies with high quality Tully-Fisher parameters (i.e. apparent magnitudes and rotational velocities), to compare the peculiar velocity predicted from the distribution of IRAS galaxies to that observed for individual objects, using a maximum likelihood method that takes into consideration the effects of an inhomogeneous mass distribution and a variety of sampling selection effects (Willick 1994). Nusser and Davis (1994) used an early version of this Mark3 dataset processed through the POTENT algorithm to compare the inferred dipole of the radial peculiar velocity field to that expected from the 1.2 Jy IRAS redshift survey data (Fisher *et al.* 1994a). They derived a best estimate $\beta \approx 0.6$ but they did not attempt to estimate the uncertainty because they had no control over the POTENT field.

Direct estimation of the peculiar velocity of individual galaxies based on the forward Tully-Fisher relationship can be used to generate a peculiar velocity field, with nice visual images, but is difficult to correct for the myriad selection biases. Less problematic in terms of Malmquist corrections is to use the inverse Tully Fisher relationship (Schechter 1980; Aaronson et al. 1982; Lynden-Bell 1991), (i.e. $\eta = \eta(M)$), but this relationship is difficult to visualize, and is best done by means of a parameterized model. Aaronson et al. (1982) fitted a spherically symmetric Virgocentric flow model to their data using an inverse TF relationship and Lahav (1991) fitted a pure bulk flow model. Roth (1993) and Willick et al. (1994b) fit a series of IRAS predicted flow models (Strauss and Davis 1988; Yahil 1988) to their data where $\beta$ used to construct the IRAS flow models is the only free parameter. Although this method can derive a 'best' estimate of $\beta$, it does not immediately tell us if the model is a good fit to the data. This procedure also suffers if the redshift-distance relation is multivalued, as frequently occurs toward clusters of galaxies.

In this paper we shall demonstrate a middle ground, in which, without binning the data, we effectively generate a smoothed version of the peculiar velocity field by means of an inverse fit to the TF relationship. A smoothed fit to the observed data has the advantage that it filters out small scale peculiar velocities that are beyond the resolution of the IRAS gravity maps, and which are not expected to obey linear theory in any event. The result is that, without ever binning the peculiar velocity data, we can generate maps of the peculiar velocity field, which we demonstrate are an unbiased estimate of the true field.

In section 2, we describe the method, its biases, and the construction of the orthogonal functions. In section 3, we construct a Monte-Carlo catalog derived from an N-body simulation to test the method, and demonstrate that the predicted velocity field is unbiased for different choices of resolution. A discussion of the utility of the technique is given in section 4. We shall apply this method to the Mark3 data in a separate publication.

## 2. THE METHOD

### 2.1 Maximum Likelihood fit for $\eta$

The goal is to derive a spatially smoothed estimate of the peculiar velocities of a sample of galaxies having measured circular velocity parameters, $\eta_i$, apparent magnitudes, $m_i$, and redshifts, $z_i$. We assume that the circular velocity parameter, $\eta$, of a galaxy is related to its absolute magnitude, $M$, by means of a linear inverse Tully-Fisher (ITF) relation, i.e.,

$$\eta = sM + \eta_0 + R, \tag{1}$$

where $\eta_0$ and $s$ are constants and $R$ is a Gaussian random component with mean 0 and $rms$ value $\sigma_\eta$ independent of $\eta$ and $M$. If $u$ is the radial peculiar velocity of a galaxy then the apparent and absolute magnitudes are related by,

$$M = m - 5\log(z - u) - 15. \tag{2}$$

where the peculiar velocity and the redshift have units of km/s. The relations (1) and (2) suggest the following procedure for estimating the peculiar velocities:

( i) Express the peculiar velocities in terms of a parameterized model.

( ii) Use the peculiar velocity model to infer absolute magnitudes, $\tilde{M}_i\,(m_i, z_i, \text{model})$, of galaxies from their apparent magnitudes and redshifts given the relation (2).

( iii) Assume Gaussian deviations from the ITF relation so that the probability of measuring the set of observed values $\eta_i$, given (1) and the model predicted absolute magnitudes, $\tilde{M}_i$, given from ( ii), is:

$$\Pr(\eta|\tilde{M}) \propto \exp\left[-\sum_{i=1}^{N_g} \frac{\left(\eta_0 + s\tilde{M}_i - \eta_i\right)^2}{2\sigma_\eta^2}\right], \qquad (3)$$

where $N_g$ is the number of galaxies in the sample.

( iv) Find the model best fit parameters by maximizing the probability function (3).

It should be noted that a peculiar velocity model in which the number of fit parameters is significantly less than the number of galaxies in the sample is fully consistent with the gravitational instability process, where large scale, coherent flows are expected to have substantial amplitude.

The procedure for finding the best fit parameters for the model is equivalent to minimizing the $\chi^2$ function,

$$\chi^2 = \sum_i \frac{\left(\eta_0 + s\tilde{M}_i - \eta_i\right)^2}{\sigma_\eta^2}, \qquad (4)$$

where $\sigma_\eta$ is constant. We now proceed to construct a form for the fit model. Using (2) we write,

$$M_i = M_{0i} + P_i \qquad (5)$$

where

$$M_{0i} = m - 5\log(z_i) - 15 \quad \text{and} \quad P_i = -5\log\left(1 - \frac{u_i}{z_i}\right). \qquad (6)$$

Since the absolute magnitudes depend linearly on the quantities $P_i$, we choose to express $P_i$ in terms of linear combination of a set of independent functions. It is the parameters of this combination which we shall determine by minimizing the $\chi^2$. Thus we write for each galaxy,

$$P_i = \sum_{j=0}^{j_{max}} \tilde{a}^j \tilde{F}_i^j \qquad (7)$$

where $j_{max}$ is the total number of the fit parameters $\tilde{a}^j$ and the matrix $\tilde{F}$ satisfies the following orthonormality condition,

$$\sum_{i=1}^{N_g} \tilde{F}_i^j \tilde{F}_i^{j'} = \delta_K^{j,j'}, \qquad (8)$$

where $\delta_K$ is the Kronecker delta and we set $\tilde{F}_i^0 = 1/\sqrt{N_g}$. After substituting (5), (6) and (7) in (4) we find,

$$\chi^2 = \frac{1}{\sigma_\eta^2} \sum_i (\eta_0 + sM_{0i} - \eta_i)^2 + \frac{2s}{\sigma_\eta^2} \sum_{i,j} (\eta_0 + sM_{0i} - \eta_i)\tilde{a}^j \tilde{F}_i^j + \frac{s^2}{\sigma_\eta^2} \sum_j (\tilde{a}^j)^2. \quad (9)$$

Thanks to the orthogonality condition (8) there is no coupling between the different $\tilde{a}^j$ modes. The TF parameters $\eta_0$ and $s$ depend on the velocity model, therefore the $\chi^2$ should be minimized subject to variations in $s$ and $\eta_0$ as well as in $\tilde{a}^j$. At the minimum, the derivatives $\partial \chi^2 / \partial \tilde{a}^j$, $\partial \chi^2 / \partial \eta_0$ and $\partial \chi^2 / \partial s$ vanish, yielding respectively,

$$\begin{aligned}
\sum_i (\eta_0 + sM_{0i} - \eta_i)\tilde{F}_i^j + s\tilde{a}^j &= 0, \\
\sum_i (\eta_0 + sM_{0i} - \eta_i) + s\sum_{i,j} \tilde{a}^j \tilde{F}_i^j &= 0, \\
\sum_i (\eta_0 + sM_{0i} - \eta_i)M_{0i} + \sum_{i,j} (\eta_0 + 2sM_{0i} - \eta_i)\tilde{a}^j \tilde{F}_i^j + s\sum_j (\tilde{a}^j)^2 &= 0.
\end{aligned} \quad (10)$$

The first two of Eqs (10) are degenerate for $j = 0$. Then the determination of $\tilde{a}^0$ is somewhat arbitrary. The reason for this ambiguity is that $\tilde{a}^0 \neq 0$ amounts to a radial peculiar velocity $u \propto z$ which is indistinguishable from the Hubble expansion. For a realistic database with finite surveyed volume about us, a non vanishing $\tilde{a}^0$ implies that the surveyed volume is a region of constant density contrast. This is highly improbable for most current observations which typically extend out to distances larger 8000 km/s. Therefore we set $\tilde{a}^0 = 0$. With this choice for the zeroth mode, the solution for $\tilde{a}^j$, $\eta_0$ and $s$ can be written in a closed form as,

$$\begin{aligned}
\tilde{a}^j &= -\frac{1}{s} \sum_i (sM_{0i} - \eta_i) \tilde{F}_i^j, \\
\eta_0 &= -\frac{1}{N_g} \sum_i (sM_{0i} - \eta_i), \\
s &= \frac{\sum_{i,i'} M_{0i}\eta_{i'} \left(1 + N_g \sum_j \tilde{F}_i^j \tilde{F}_{i'}^j\right) - N_g \sum_i M_{0i}\eta_i}{\sum_{i,i'} M_{0i}M_{0i'} \left(1 + N_g \sum_j \tilde{F}_i^j \tilde{F}_{i'}^j\right) - N_g \sum_i M_{0i}^2}.
\end{aligned} \quad (11)$$

The terms involving the matrix $\tilde{F}$ are the correction in the estimate of the ITF parameter $s$ arising from the peculiar velocities. The covariance matrix can be computed rigorously. However if the correction terms are small, we may neglect the coupling between $\tilde{a}^j$ and $s$. There is no coupling between $\tilde{a}^j$ and $\eta_0$ because of the orthogonality of the modes and our choice of the zeroth mode. This implies that for Gaussian scatter of $\eta$ around the ITF, the errors are normal and given by,

$$\sigma_{\tilde{a}^j} = \left(\frac{1}{2}\frac{\partial^2 \chi^2}{\partial(\tilde{a}^j)^2}\right)^{-1/2} = \frac{\sigma_\eta}{s},$$

$$\sigma_{\eta_0} = \left(\frac{1}{2}\frac{\partial^2 \chi^2}{\partial \eta_0^2}\right)^{-1/2} = \frac{\sigma_\eta}{\sqrt{N_g}}, \qquad (12)$$

$$\sigma_s = \left(\frac{1}{2}\frac{\partial^2 \chi^2}{\partial s^2}\right)^{-1/2} = \sigma_\eta \left(\sum_i M_{0i}^2\right)^{-1/2}.$$

It is not surprising that the errors in $\tilde{a}^j$ scale like $1/s$. Indeed, for $s = 0$ the ITF is superfluous and cannot constrain the peculiar velocity field. Using the linear combination (7) we find that the error in the determination of the quantity $P_i$ of any individual galaxy is,

$$\sigma_{P_i} = \frac{\sigma_\eta}{s}\left(\sum_{j=1}^{j_{max}} (\tilde{F}_i^j)^2\right)^{1/2} \qquad (13)$$

The orthonormality condition (8) implies that $\sigma_{P_i} \propto \sqrt{j_{max}/N_g}$. The corresponding error in the velocity in the limit $u/z \ll 1$ is,

$$\sigma_{u_i} = 0.46\ \sigma_{P_i} z_i \qquad (14)$$

In the limit $j_{max} \to N_g$ we recover the expected error, $\sigma_{u_i} \approx \sigma_\eta z_i/s$.

### 2.2 Biases in The Method

The ITF method described above is intended to provide a generalized smoothed fit to the peculiar velocity data, and is not in itself motivated by any particular assumptions about the nature of the flow. We do not, for example, require the flow to be irrotational, as is required for POTENT. The fit makes no assumptions about the conversion of redshift space to real space, and simply seeks the best smoothed fit in redshift space. An important question to ask of any method is whether it suffers from any systematic biases, and if so, can they be readily calibrated and corrected.

There are three effects which bias the estimation of peculiar velocities from the ITF. First, there might be some selection effects associated with measuring circular velocities. An example of this is the reduced signal to noise ratio of broad 21cm lines, relative to narrow lines of the same equivalent width. This is not considered to be a problem in practice, as galaxies with larger rotational width usually contain more hydrogen than those with slower rotation. Furthermore, many of the recent observations of line width are based on $H_\alpha$ rotation curves derived from long slit spectra, for which there is no signal to noise penalty for broad lines.

A second, more important effect, is that small scale velocity dispersion introduces homogeneous as well as inhomogeneous Malmquist bias (e.g. Lynden-Bell 1991, Willick 1994). To see the effect of velocity dispersion, consider the following example. Suppose that we observe a set of $(\eta_i, m_i)$ for $N_g$ galaxies, all at the same redshift, $z$, and direction in the sky. Assume that the parameters $\eta_0$ and $s$ of the ITF have already been determined. Let us for the sake of simplicity consider a case in which all the galaxies have a common radial peculiar velocity component, uncorrelated random components, $\delta u_i$, with 0 mean. We seek an estimate, $\bar{u}_{est}$, for $\bar{u}_{true}$. By maximizing the likelihood for measuring the set of $\eta_i$ with respect to the estimate, $\bar{u}_{est}$, we find,

$$5sN_g \log\left(1 - \frac{\bar{u}_{est}}{z}\right) = \sum_{i=1}^{N_g}(\eta_0 + sM_{0i} - \eta_i) \tag{15}$$

But $M_{0i} = M_i + 5\log(1 - u_i/z)$, where $u_i = \bar{u}_{true} + \delta u_i$ are the peculiar velocities of the galaxies. Since $\sum_i(\eta_0 + sM_i - \eta_i) = 0$, we obtain,

$$N_g \log\left(1 - \frac{\bar{u}_{est}}{z}\right) = \sum_i \log\left(1 - \frac{u_i}{z}\right) \tag{16}$$

In the absence of small scale peculiar velocities, we have $\delta u_i = 0$ and $u_i = \bar{u}_{true}$. Thus the last equation yields an unbiased $\bar{u}_{est} = \bar{u}_{true}$. However, if $\delta u_i \neq 0$, then, by expanding the logarithmic terms to first order, we find,

$$\bar{u}_{est} = \bar{u}_{true} + \frac{1}{N_g} \sum_{i=1}^{N_g} \delta u_i \tag{17}$$

The sum of $\delta u_i$ evaluated at a given redshift does not necessarily vanish on the average, and the resulting bias $\Delta u_{bias} \equiv \sum \delta u_i / N_g$ caused by the velocity dispersion can be written in the general form,

$$\Delta u_{bias} = \frac{\int \delta u \Pr(\delta u) \Pr(r = z - u) du}{\int \Pr(\delta u) \Pr(r = z - u) du}, \tag{18}$$

where we assume that the probability distribution of $\delta u$ is independent of position and that $\Pr(r)$, the probability that an object lies at a position $r$, is $r^2 n(r)\Phi(r)$ where $n(r)$ is the underlying number density and $\Phi(r)$ represents the selection probability that an object is actually identified in the sample. If $\delta u$ is normally distributed about $\bar{u}_{true}$ with dispersion $\sigma_u$, then, by expanding $\Pr(r = z - u)$ to first order in $u$ about $z$, Eq(18) yields,

$$\Delta u_{bias} = \sigma_u^2 \frac{d\ln[z^2 n(z)\Phi(z)]}{z\, d\ln z}. \tag{19}$$

Approximating the selection function in a finite region by a power law $\Phi \propto r^{-\alpha}$, we find,

$$\Delta u_{bias} = \left(2 - \alpha + \frac{d\ln n}{d\ln z}\right) \frac{\sigma_u^2}{z}. \tag{20}$$

For typical values, $\sigma_u = 200$ km/s, $z = 2000$ km/s, the bias is 40 km/s for a homogeneous distribution of objects in a volume limited catalog ($n = const$, $\alpha = 0$). An inhomogeneous spatial distribution of galaxies, however, would be expected to lead to larger bias in the components of the fitted flow.

We point out that this same bias occurs with a larger effective value of $\sigma_u$ in multi-valued zones, where objects with very different distances have similar redshifts, due to infall peculiar velocities toward a high density peak. Again, the bias is expected to depend inversely on the redshift. However the dependence on the density run and the selection function is more difficult to evaluate as the distribution of peculiar velocities in a multi-valued zone depends on the profile of the density peak. In realistic applications, an assessment of this bias in the ITF estimated peculiar velocities may be made by inspecting the multi-valued zones seen in the prediction of velocity fields from galaxy redshift surveys such as the 1.2Jy IRAS.

In our methodology, the smoothed velocity field is deduced from the modeled function $P$. This is desirable since it avoids a possible bias in the calibration of the ITF parameters within magnitude limited surveys due to small scale velocity dispersion. Indeed, in magnitude limited surveys, objects at larger redshifts tend to have larger absolute luminosity. To second order in $u/z$, $P \approx 2.17(u/z + (u/z)^2/2)$. If the velocity field at redshift $z$ is $\bar{u}$, then averaging over a sample of galaxies with small scale dispersion $\sigma_u$ at this position gives an average $P$ of

$$\bar{P} \approx 2.17 \left( \frac{\bar{u}}{z} + \frac{\bar{u}^2}{2z^2} + \frac{\sigma_u^2}{2z^2} \right) \quad .$$

The first two terms represent the estimate of $\bar{P}$ in terms of the linear modeling of the peculiar velocity field, but the third term amounts to a bias that diminishes with distance, i.e. with absolute luminosity. Therefore the inferred slope of the ITF relation is expected to be biased by the presence of small scale velocity noise when expressing velocity field in a linear model. The bias would result in a flatter slope $s$ if $\bar{u}$ is the true mean peculiar velocity. Since the model velocity field and ITF parameters are found self-consistently, this bias would affect the estimated peculiar velocities. However, by using a linear model for $P$ rather than $u$, we avoid this form of calibration bias, but at the same time, a third bias of our estimate of the velocity is introduced. Our estimate of the peculiar velocity at a redshift $z$ can, to second order in $P$, be written as,

$$\frac{\bar{u}_{est}}{z} = 0.46 \left( \bar{P} - 0.23 \bar{P}^2 \right).$$

But the true mean peculiar velocity is obtained by averaging the quantities $P_i$ of all objects at the redshift $z$. Therefore the true mean peculiar velocity is,

$$\frac{\bar{u}_{true}}{z} = 0.46 \left( \bar{P} - 0.23 \bar{P}^2 - 0.23 \sigma_P^2 \right),$$

where $\sigma_P$ is the *rms* dispersion of $P$ about $\bar{P}$ due to small scale velocity dispersion. And, for simplicity, we have ignored the bias in $\bar{P}$ caused by averaging at the redshift and not the true distance of objects. Hence our estimate of $\bar{u}_{est}$ is biased by the presence of the term with $\sigma_P$. To the relevant order, $\sigma_P \propto \sigma_u/z$, hence the bias is $\propto \sigma_u^2/z$ which is also the behavior of the bias (20). At large redshifts this bias is not important since there $u \propto P$ and $\bar u \propto \bar P$. In our method the slope of the ITF as inferred from the solution (11) differs from the true one only because of the bias given by (20).

Clearly, the biases associated with the small scale velocity dispersion discussed here can be avoided by restricting the application of the method to large enough redshifts.

### 2.3 Construction of The Orthonormal Matrix

To implement the method we need to construct the matrix $\tilde F$. In principle this can be done by performing a Gramm-Schmidt algorithm on any set of independent functions. In the case of a uniform distribution of matter, an orthogonal base of functions is formed from the radial spherical Bessel and the angular spherical harmonic functions (Regös & Szalay 1989, Fisher *et al.* 1994b). Here we shall use Bessel functions to describe the radial behavior of the angular multipoles obtained by the expansion in spherical harmonics. We write the quantity $P$ in the form,

$$P(z,\theta,\phi) = \sum_{n=0}^{n_{max}} \sum_{l=0}^{l_{max}} \sum_{m=-l}^{m=l} \frac{a_{nlm}}{z} \left(j'_l(k_n z) - c_{l1}\right) Y_{lm}(\theta,\phi) \qquad (21)$$

where $j'_l$ is the derivative of the spherical Bessel of order $l$, $Y_{lm}$ are the angular spherical harmonics, and $a_{nlm}$ is the amplitude of the mode with $l, m$ and $n$. The factor $c_{l1}$ is non-zero only for $l = 1$. The actual value of $c_{l1}$ is determined by the reference frame relative to which the redshifts are measured and hence the peculiar velocities are estimated. The natural choice is to work in a frame in which the observer is at rest relative to the flow of matter near the origin (e.g. Nusser & Davis 1994). In the real universe this would be the Local Group frame. In such a frame, $P$ vanishes near the origin, hence $c_{11} = \lim_{x \to 0} j'_1(x)$. In the limit $|u/z| \ll 1$, which is expected both near the origin and at the outer limit, we have $P \propto u/z$. Therefore the expansion (21) is well motivated because it corresponds to the correct asymptotic behavior at the origin of a velocity potential $\Phi$, $\Phi = \sum a_{lmn} j_l(k_n) Y_{lm}(\theta,\phi)$ (Jackson 1962).

Therefore we set the basis set of functions to be orthonormalized on the discrete non-uniform distribution of the data points as,

$$F_i^j = \frac{1}{z} \left(j'_l(k_n z_i) - c_{l1}\right) Y_{lm}(\theta_i, \phi_i), \qquad (22)$$

where $j = 1 \ldots j_{max}$ is defined by each triplet $n, l,$ and $m$. It is those functions which we orthonormalize using the Gramm-Schmidt algorithm. The orthonormal functions are denoted by $\tilde F$.

The expansion of an arbitrary radial function into a set of Bessel functions requires the specification of additional boundary conditions appropriate for the problem in question. These boundary conditions determine the wavenumbers $k_n$ in the expansion (21). In any realistic application, it is necessary to set a maximum radius, $z_{max}$, within which the data is sufficiently reliable to constrain the peculiar velocity field. It is at $z_{max}$ where the outer boundary conditions are specified. We are guided by the asymptotic behavior of a peculiar velocity field generated by fluctuations in the mass-density distribution. If we assume that the distribution of matter beyond $z_{max}$ is uniform, then potential theory dictates that $u \propto z^{-(l+2)}$ at $z > z_{max}$. In addition to a decaying mode $z^{-3}$, the dipole harmonic, $l = 1$, attains a constant value which is the reflection of the observer's motion relative to the bulk motion of matter at $z \geq z_{max}$. Thus we require

$$\frac{\mathrm{d}\ln(j_l(k_n z_{max}))}{\mathrm{d}\ln(z_{max})} = -(l+2), \tag{23}$$

which, using the recurrence relations of spherical Bessel functions, leads to $j_{l-1}(k_n z_{max}) = 0$ (Fisher *et al.* 1994b). We solve for the lowest values of $k_n$ which satisfy this relation for each $l$.

Once the zeros are fixed we introduce additional quadrupole modes which take into account the velocity components generated at $z < z_{max}$ by mass fluctuations at $z > z_{max}$. An external quadrupole field generates a peculiar velocity which scales as $z$; External fluctuations do not generate dipole and monopole velocity terms at $z < z_{max}$, and we neglect higher order modes. This external quadrupole mode is not orthogonal to the oscillatory Bessel function modes, but this does not prevent including it within a complete basis set of functions. The external source terms are expected to make a negligible contribution to the predicted velocity field in the internal region unless the external density fluctuations have a gross quadrupole term.

The total number of the model fit parameters is fixed by the highest order of the spherical harmonic, $l_{max}$ and the number of radial modes in the Bessel function expansion, $n_{max}$. Including the external modes for $l = 2$, we have

$$n_{modes} = n_{max}(l_{max}^2 + 2l_{max} + 1) + 5 \tag{24}$$

valid for $l_{max} \geq 2$. Thus for a typical case, $n_{max} = 8$ and $l_{max} = 3$ we have 133 degrees of freedom in the fit.

## 3. TESTING WITH AN N-BODY SIMULATION

We test the reconstruction of peculiar velocities from the ITF using a standard CDM N-body simulation normalized to have $rms$ mass fluctuations $\frac{\delta\rho}{\rho} = 0.7$ in a sphere of $8h^{-1}$Mpc. Particle velocities in the simulation were smoothed with a Gaussian window of width $1h^{-1}$Mpc so that the pair velocity dispersion on a scale of $1h^{-1}$Mpc is close to the observed value, 317km/s (Fisher *et al.* 1994a). A particle moving at 600 km/s, with small local shear, and not in a dense region, was chosen as the central observer (Fisher *et al.* 1994a). The coordinates were rotated so that this observer is moving in the same direction of the sky as our Local Group of Galaxies, according to the CMBR dipole anisotropy. A mock full sky 1.2Jy IRAS catalog was generated from the simulation by assigning luminosities to particles according to the luminosity function of the 1.2Jy IRAS survey. This procedure resulted in 4881 "galaxies" out to a distance of $180h^{-1}$ from the "observer".

To further improve the realism of the simulation, we scanned the galaxy positions of the Mark3 catalog (Willick *et al.* 1994a), which is quite anisotropically selected across the sky, and selected the nearest simulation "galaxy" to each Mark3 real galaxy, resulting in a final catalog of 2871 "galaxies". An inverse TF relation with $\eta_0 = -2$ and $s = -0.1$ was used to generate three mock Mark3 catalogs with $\sigma_\eta = 0.01, 0.03$ and $0.07$.

In Figure 1 we show the reconstruction of the peculiar velocities from the ITF relation using the fake circular velocities and apparent magnitudes of the 4881 "galaxies" in the mock IRAS catalog with $\sigma_\eta = 0.01$. The reconstruction method was applied once with $(n_{max}, l_{max}) = (8, 3)$ and once with $(n_{max}, l_{max}) = (12, 4)$ radial and angular modes respectively. The maximum redshift at which the external boundary conditions used to determine the zeros of the Bessel functions was $z_{max} = 10000$ km/s. We see that the method is successful at recovering the true velocities of "galaxies" in the simulation. The reconstruction with $(n_{max}, l_{max}) = (12, 4)$ shows more scatter than the one with $(n_{max}, l_{max}) = (8, 3)$. The reason is that the errors in the estimated velocities increase with the total number of fit parameters. In other words, though the $\chi^2$ improves as the number of fit parameters is increased, it attains a wider minimum. The difference between the estimated and the true peculiar velocities as a function of redshift is shown in Figure 2. As expected the scatter increases with redshift. Yet, no biases are detected in the estimated versus true velocities at any redshift. The predicted slope of the ITF relation as given from the solution (11) is $s = -0.098$ while the true value is $s = -0.1$, demonstrating that this bias is negligible.

Most of the scatter seen in Figures 1 and 2 is a result of the small scale velocity dispersion which is not fitted by the smoothed field, since the scatter in the ITF, $\sigma_\eta = 0.01$, translates to a magnitude error of $\sigma_M = \sigma_\eta/s = 0.1$, or a peculiar velocity error of $0.05z$, considerably smaller than is seen in Figure 2.

Figure 3 shows the reconstruction of the true peculiar velocities of the mock Mark3 galaxies for the lower resolution smoothing. Only velocities of "galaxies" within a redshift of 5000 km/s are plotted. The two upper and the lower right panels of Figure 3 show the reconstruction of the true peculiar velocities of the mock Mark3 data for the three values of $\sigma_\eta$. Most encouraging is that no systematic bias in the estimate of the peculiar velocities is evident. The scatter in the plots for $\sigma_\eta = 0.03$ and $0.07$ is attributed to the scatter in the ITF relation and the small scale intrinsic velocity dispersion. The predicted slopes of the ITF are $s = -0.0978, -0.0974$ and $-0.0964$ for $\sigma_\eta = 0.01, 0.03$ and $0.07$. To measure how well our smoothed estimate of the peculiar velocities can match the velocity field as predicted from the redshift space distribution of galaxies, we show in the lower left panel the smoothed peculiar velocities with $\sigma_\eta = 0.03$ against the linear theory velocities predicted from the density field in the mock IRAS catalog. The velocities from the density field were obtained using the linear theory method of Nusser & Davis (1994). It seems that the reconstruction of the velocities from the simulated ITF and the density field agree very well, with no apparent bias in the slope of the correlation. This result motivates the application of this smoothing method to real data so as to better constrain the value of $\beta = \Omega^{0.6}/b$. The difference between the $Y$ and $X$ axes of each panel in Figure 3 is shown versus redshift in Figure 4. Again we note that even in sparse data like the Mark3 catalog, the method is capable of recovering the peculiar velocities without any noticeable biases.

Figure 5 shows the ratio of the velocity differences from Figure 4 to the error in the estimated velocities versus redshift. The errors were approximated using Eq(13). Since $\sigma_u \propto z$, the scatter decreases with redshift. In the absence of any small scale velocity dispersion, the *rms* value of the scatter in this plot would be unity. The large scatter for $\sigma_\eta = 0.01$ indicates that small scale velocities amounts to most of the random error in the estimate of peculiar velocities of individual "galaxies". But the scatter caused by small scale velocity dispersion is not negligible even for $\sigma_\eta$ as large as 0.07. No systematic bias of the estimated relative to the true peculiar velocities is evident. However, the ratio corresponding to the linear theory prediction plotted in the lower left panel shows some positive offset.

In Figure 6a and 6b we examine in more detail the nature of the agreement between the smoothed and the true peculiar velocities of the mock Mark3 "galaxies". with $\sigma_\eta = 0.03$. We show a full sky plot of the positions of all galaxies with $3500 < z < 5000$ km/s. The differences between the estimated and true velocities are in very good agreement considering that the estimated velocities are smoothed on a scale $\sim 1000$ km/s.

In order to demonstrate the possible biases which might be present in estimated peculiar velocities, we have boosted the small scale velocity dispersion by assigning to "galaxies" in the mock IRAS catalog additional radial peculiar velocity components drawn from a normal distribution with dispersion $\sigma_u = 400$km/s so that the total small scale velocity dispersion is $\sim 500$km/s. The new peculiar velocities were then used to compute the redshifts of "galaxies". In Figure 7 we plot the difference between the estimated and true peculiar velocities versus redshift. We used $\sigma_\eta = 0.01$ in this reconstruction. It is clear the the estimated velocities are biased. But even for such a large value of the velocity dispersion the bias is not severe. The sense of the bias is that at large and low redshifts ($z > 4000$km/s, $z < 2000$km/s) the method underestimates the true peculiar velocities. The bias at large redshifts is caused by the lower slope predicted by the method. The predicted slope using the solution (11) is $s = -0.092$ while the true value is $s = -0.1$.

As an indication of the coherence of the residuals of the fit $\Delta u_i$, we display in Figure 8 the correlation function of the velocity residuals, computed in the same fashion as Gorski *et al.* (1989). We plot the quantity

$$\Psi_v(\Delta s) = \frac{\sum_{i,j} W_i W_j \Delta u_i \, \Delta u_j \cos(\theta_{ij})}{\sum_{i,j} W_i W_j \cos^2(\theta_{ij})} \qquad (25)$$

where the sum is over all pairs in the mock Mark3 catalog ($\sigma_\eta = 0.03$) separated by redshift separation $\Delta s$, $\cos(\theta_{ij})$ is the angle on the sky between the galaxies $i$ and $j$, and the weight function is arbitrary but is here chosen to be unity, $W_i = 1$ for $z < 6000$ km/s and $W_i = 0$ for $z > 6000$ km/s. The short dashed curve in Figure 8 shows $|\Psi_v|^{1/2}$ for the true peculiar velocity of the mock Mark3 data, not the residuals; in the local group frame shown here, the field is almost completely coherent. The solid line shows $|\Psi_v|^{1/2}$ (with sign preserved) of the fitted minus true peculiar velocity for the Mark3 mock sample, while the dashed curve shows $|\Psi_v|^{1/2}$ for the fitted minus linear theory peculiar velocity. Note in the latter case that the coherence of the residuals is considerably larger than for the fit-true velocities. This coherence is consistent with the small offset of residuals from zero seen in the lower left panel of figure 5 (compare to upper right panel of figure 5) and is due to the nonlocal nature of the velocity field inferred from linear theory. Errors in this field due to shot noise will result in coherent velocity residuals. The extra coherence exhibited in Figure 5 and Figure 8 for the fitted minus linear theory fits is consistent with previous estimates of the reconstruction errors for IRAS type gravity maps (Strauss *et al.* 1992). The flat behavior of $\Psi_v$ is indicative of an error in the overall dipole amplitude of the IRAS-like fit, of order 100 km/s,

It is instructive to examine the value of $\chi^2$ versus the number of modes included in the fit, since this allows an assessment of the statistically significant modes used in the reconstruction. The orthogonal function method is very useful for this purpose. The fit parameters $\tilde{a}^j$ can easily be arranged in descending order, $\tilde{a}^{j_1} \geq \tilde{a}^{j_2} \ldots \geq \tilde{a}^{j_{max}}$, and $\chi^2$ can be computed as a function of the limiting mode. Figure 9 shows $\chi^2$ versus the number of modes incorporated in the reconstruction, for the case $(n_{max}, l_{max}) = (8, 3)$ in the mock Mark3 catalog constructed with $\sigma_\eta = 0.03$. We see that for a number of modes higher than $j = 100$ the gain in the $\chi^2$ per degree of freedom is marginal. The "saturation" of $\chi^2$ as a function of degrees of freedom is the criterion according to which the statistically significant fit parameters are determined in realistic applications. Note that, due to the small scale velocity dispersion which is not fitted by the model, the $\chi^2$ of the model per degree of freedom is larger than unity, even when all the modes are included.

## 4. DISCUSSION

We have presented a method for estimating the peculiar velocities of disklike galaxies given measurements of their apparent magnitudes and circular velocity parameters. The method depends on a linear approximation of the inverse Tully-Fisher relationship and is free of Malmquist bias in the absence of multi-streaming. The bias induced by the observed amplitude of small scale peculiar velocities is essentially negligible. In any case the bias can be completely avoided by discarding nearby galaxies in the estimation of peculiar velocities. The peculiar velocities are given by means of an expansion in a set of independent and orthogonal base functions. Therefore the method successfully recovers the smoothed components of the peculiar velocities of galaxies without the necessity of prior binning of the data. The choice of the base functions determines the form of the smoothing employed. For example, in the description of a discrete set of radial Bessel and angular spherical harmonics, which correspond to an expansion of a 3D Fourier series, the smoothing is equivalent to a sharp cutoff in $k-$space. A general window function may be introduced by multiplying the Bessel and spherical harmonics expansion by its inverse. The model peculiar velocity field can be used to generate smoothed maps of the peculiar velocity field in regions of space where data is not available, although the inversion of the matrix of the orthogonal functions is singular and will require special care. The choice of the window function clearly depends on the goal of the application of the method. For instance, in order to compare the inverse Tully-Fisher estimated velocities with the IRAS predictions, a variable smoothing is appropriate. On the other hand if we seek peculiar velocity maps for comparisons with the predictions of cosmological scenarios, a large constant smoothing is desired.

The asymptotic flatness of the $\chi^2$ curve (Figure 9) at a level above that expected for a "good" fit to the data is one indication that our particular choice of basis functions is not ideally matched to the problem at hand; too many oscillations of the spherical Bessel functions occur at large distance where the constraints are few. A larger number of basis functions only slightly increases the parameters available for better fitting the local flow, while introducing still more unconstrained oscillations at larger distances. A better optimized set of basis functions should allow us to reach the same or better $\chi^2$ value with fewer fit parameters.

The method presented here is designed to give a best estimate for the peculiar velocities of galaxies by means of their redshifts. Therefore the associated smoothed peculiar velocity field is given in redshift space rather than real space, which is ideal for comparison to the peculiar velocity field computed by Nusser and Davis (1994), which was also in redshift space. A somewhat cumbersome property of flows in redshift space is that the velocity field is not vorticity free when the flow is non-linear. This poses some complication when seeking a recovery of the mass-density fluctuations, from the velocity field (e.g. Nusser *et al.* 1991). Therefore it is worthwhile to obtain an estimate for the velocity field in real space. The method proposed here can easily be modified to iteratively solve for the velocity field in real space. In a separate paper we intend to test the recovery of the velocity field in real as well as redshift space using various smoothing window functions which are of interest in many applications of the method. The comparison of the smoothed velocities derived from the actual Mark3 data with the 1.2Jy IRAS predicted velocities is underway and shall be reported in due course.

### Acknowledgments

We thank Karl Fisher and Ofer Lahav for useful conversations. This work was supported in part by NSF grant AST 92-21540 and NASA grant NAG 51360. AN acknowledges the support of a SERC postdoctoral fellowship.

# FIGURE CAPTIONS

**Figure 1 :** The estimated peculiar velocities using the inverse Tully-Fisher relationship *vs.* the true peculiar velocities for galaxies in the mock IRAS catalog. Only velocities of "galaxies" within $z = 5000$km/s are plotted. The left panel shows the reconstruction for $(n_{max}, l_{max}) = (12, 4)$ while the right panel is for the reconstruction with $(n_{max}, l_{max}) = (8, 3)$. Velocities are in units of km/s.

**Figure 2 :** The difference between the estimated and true peculiar velocities *vs.* redshift of the mock IRAS "galaxies" in figure 1. Redshift is in units of km/s.

**Figure 3 :** The estimated *vs.* true peculiar velocities of "galaxies" in the mock Mark3 data for three values of $\sigma_\eta$ (upper and lower right panels). Also shown (lower left panel) the estimated vs. peculiar velocities of the mock Mark3 "galaxies" *vs.* the prediction of peculiar velocities from the density fluctuations in the mock IRAS survey. Only velocities of "galaxies" within $z = 5000$ km/s are plotted.

**Figure 4 :** The difference between the estimated and N-body peculiar velocities from figure 3 *vs.* redshift.

**Figure 5 :** The ratio of the velocity differences from figure 4 to the $1\sigma$ error in the estimated velocities *vs.* redshift. The errors were computed using equation (13).

**Figure 6 :** The estimated and true peculiar velocities in the mock Mark3 "galaxies" on a spherical shell between $z = 3500$ and $5000$ km/s. Open circles denote galaxies infalling to us (LG frame); stars are outflowing from us. The size of the symbol is proportional to the amplitude of the flow; a flow velocity of 200 km/s is indicated in the lower right.

**Figure 7 :** The difference between the estimated and true peculiar velocities *vs.* redshift of the mock IRAS "galaxies". In addition to their true peculiar velocities, the "galaxies" were assigned random radial peculiar velocities with dispersion of $\sigma_u = 400$ km/s.

**Figure 8 :** The square root of the correlation function of velocity residuals (units (km/s) versus pair separation. Zero lag corresponds to the *rms* amplitude of the residuals. The solid curve is the correlation of the fitted minus true N-body velocity, while the long dashed curve is the correlation function of the fitted minus linear theory velocity. The covariance of the true velocity field is shown as the short dashed curve.

**Figure 9 :** The $\chi^2$ *vs.* the number of modes incorporated in the reconstruction of the model. The fit parameters were first put in descending order.